\documentclass{ecai}
\usepackage{threeparttable}
\usepackage{latexsym}
\usepackage{hyperref}
\usepackage{algorithm}
\usepackage{algorithmic}
\usepackage[numbers,sort&compress]{natbib}
\usepackage{graphicx}
\usepackage{tabularx}
\usepackage{array}
\usepackage{float}    
\usepackage{booktabs}
\usepackage{multirow}
\usepackage{makecell}
\usepackage{colortbl}
\usepackage[table]{xcolor} 
\definecolor{lightgray}{gray}{0.3}
\usepackage{threeparttable}
\usepackage{bm}

\paperid{641}
\begin{document}

\begin{frontmatter}

\title{Tab-Attention: Self-Attention-based Stacked Generalization for Imbalanced Credit Default Prediction}

\author[A]{\fnms{Yandan}~\snm{Tan} \orcid{0000-0002-9629-1118}}
\author[B]{\fnms{Hongbin}~\snm{Zhu}\thanks{Corresponding Author. Email: zhuhb@fudan.edu.cn}}
\author[A,B]{\fnms{Jie}~\snm{Wu}}
\author[A,B]{\fnms{Hongfeng}~\snm{Chai}}

\address[A]{School of Computer Science, Fudan University, Shanghai, China}
\address[B]{Institute of FinTech, Fudan University, Shanghai, China}

\begin{abstract}
Accurately credit default prediction faces challenges due to imbalanced data and low correlation between features and labels. Existing default prediction studies on the basis of gradient boosting decision trees (GBDT), deep learning techniques, and feature selection strategies can have varying degrees of success depending on the specific task. Motivated by this, we propose Tab-Attention, a novel self-attention-based stacked generalization method for credit default prediction. This approach ensembles the potential proprietary knowledge contributions from multi-view feature spaces, to cope with low feature correlation and imbalance. We organize multi-view feature spaces according to the latent linear or nonlinear strengths between features and labels. Meanwhile, the $f1$ score assists the model in imbalance training to find the optimal state for identifying minority default samples. Our Tab-Attention achieves superior $Recall_1$ and $f1_1$ of default intention recognition than existing GBDT-based models and advanced deep learning by about 32.92$\%$ and 16.05$\%$ on average, respectively, while maintaining outstanding overall performance and prediction performance for non-default samples. The proposed method could ensemble essential knowledge through the self-attention mechanism, which is of great significance for a more robust future prediction system.

\end{abstract}

\end{frontmatter}

\section{Introduction}
%

Credit risk is a critical aspect of financial risks, where credit default is a major manifestation. Credit defaults pose a significant impact on financial market stability and economic well-being \cite{RN105}. Rising defaults have been observed globally, due to slowed economic growth, escalating geopolitical risks, and trade frictions \cite{RN106}. These defaults have far-reaching consequences, including sluggish economic growth, diminished investor confidence, and substantial losses for financial institutions. Therefore, it is imperative to develop efficient credit risk management strategies to mitigate the adverse impact of credit defaults on both the financial market and the economy.

However, credit default prediction is a challenging task due to the data imbalance, high sparsity, and weak correlation between features and labels. In recent years, recent research endeavours have primarily focused on statistical learning and machine learning techniques to mitigate the risk of credit default.

{\bf Statistical learning}: Traditionally, credit ratings play a crucial role in determining credit risk. However, the credit rating process is costly and can be influenced by subjective factors, such as the differences in expert experience and consideration standards\cite{RN80, RN81}. To overcome these issues, linear discriminant analysis (LDA) was introduced. However, LDA requires data to meet basic assumptions, such as multivariate normality and equal covariance matrices. Logistic regression (LR) is a more flexible method than traditional linear regression \cite{nie2011credit}, but it is still limited in its ability to handle complex nonlinear relationships. LR is also sensitive to outliers and missing data, and it is difficult to train LR models on large datasets. Despite these limitations, LR is a widely used method for credit default prediction due to its good interpretability. However, the global financial crisis reveals some shortcomings of LR methods, such as their slow adaptability to changing economic conditions and their limited ability to model complex nonlinear interactions among economic, financial, and credit variables.

{\bf Machine learning techniques}: The methods are crucial in adapting to changes in time and available data, while accommodating a large number of feature sets. Decision tree (DT) could capture the interactions and non-linear relationships between variables, thereby providing good discrimination between default and non-default cases \cite{nie2011credit}. However, repeated attribute segmentation for DT is susceptible to overfitting. Ensemble learning addresses this issue by growing multiple trees and computing their mean value \cite{RN90}. For example, gradient-boosting decision trees (GBDT) learn a series of weak learners to predict outputs, where weak learners are typically non-differentiable standard DTs.  Chen \textit{et al}. \cite{RN91} demonstrated that GBDT outperforms other methods in tabular data applications. Several GBDT algorithms, such as XGBoost \cite{RN87,RN95}, LightGBM \cite{RN88}, and CatBoost \cite{RN89}, have been developed to predict credit risk successfully. Although these ensemble algorithms exhibit differences, their performance is often similar across various tasks \cite{RN89}. 

Deep learning has emerged as a promising approach for credit default prediction due to its ability to capture complex patterns and nonlinear relationships between variables. Tan \textit{et al}. \cite{RN92} and Yang \textit{et al}. \cite{RN93} proposed improved DNN for superior credit default prediction than machine learning and CNN. Luo \textit{et al}. \cite{RN94} developed credit risk assessment models using deep belief networks, which excel in learning effective feature representations and capturing latent patterns in the data. Fu \textit{et al}. \cite{RN96} combined DNN and bidirectional long short-term memory to accurately identify potential credit risks. These studies demonstrate that deep learning techniques can significantly enhance the predictive performance of credit default models, making them valuable tools in credit risk management. 

Furthermore, some novel methods are based on multi-view feature organization to improve model performance. Song \textit{et al}. \cite{RN100} designed multi-view-based feature sampling to optimize imbalanced learning. Tan \textit{et al}. \cite{RN92} proposed that multi-view learning of features based on discrete, continuous, and their correlation sign differences is an effective strategy to improve prediction accuracy. In addition, some related studies have shown that feature selection based on genetic algorithms \cite{RN102}, random forest (RF) \cite{RN101}, or regularization strategies \cite{RN104} is conducive to model performance, as the feature selection process eliminates irrelevant and redundant features. 

In general, existing works have shown that GBDT \cite{RN87,RN88,RN89,RN91,RN95}, deep learning \cite{RN92,RN93,RN94,RN96}, effective feature selection strategies \cite{RN102,RN101}, and multi-view ensemble \cite{RN92,RN100} could significantly improve model performance. However, there is a limited exploration on how to combine these optimization strategies to obtain better solutions. 

For tabular data, Arik \textit{et al}. \cite{RN83} proposed an 
 attention-based TabNet, which adaptively selects less important features to apply to tabular data, resulting in improved generalization performance. Additionally, Popov \textit{et al}. \cite{RN98} designed neural oblivious decision ensembles (NODE), which combines the hierarchical decision-making process of decision trees with the representation learning ability of deep learning networks. This integration allows NODE to capture complex interactions between input features.

To comprehensively deal with the challenge of the low correlation between features and labels and data imbalance, we propose a Tab-Attention, a self-attention-based stacked generalization learning approach, for predicting credit defaults. The method mainly includes the following three key modules:

\begin{itemize}
    \item \textbf{Multi-view feature spaces}: For low correlation between features and labels in credit data, feature selection plays an important role in filtering key information. Various feature selection strategies can obtain feature sets with different advantages. Therefore, to focus on critical information and increase the informative diversity, we develop multiple feature importance-based strategies to organize multi-view feature spaces, mainly including linear correlation, nonlinear relationship, and similarity measures. Then, we design multi-layer perceptions (MLPs) to obtain outputs in different views as local-view knowledge. Besides, for all features, MLP was directly constructed to obtain global feature space. Notably, the feature sets with different importance provide key local view knowledge, which can reduce interference from less important information.

    \item \textbf{Self-attention-based stacked generalization learning}: The self-attention mechanism can dynamically weigh and combine key information. In this work, we develop a self-attention-based stacked generalization learning to efficiently ensemble information from multiple views. This approach learns ensemble mechanisms that can effectively combine the predictions of multiple models to improve the overall performance of the model.
    
    \item \textbf{Imbalanced learning based on  $f1$ score}: To tackle the issue of imbalanced data, we adopt an $f1$ score to assist in training Tab-Attention, to find the optimal state in identifying default users, so as to achieve the goal of imbalanced learning.
\end{itemize}

\section{Preliminaries}
\begin{figure}
    \centerline{\includegraphics[width=1\linewidth]{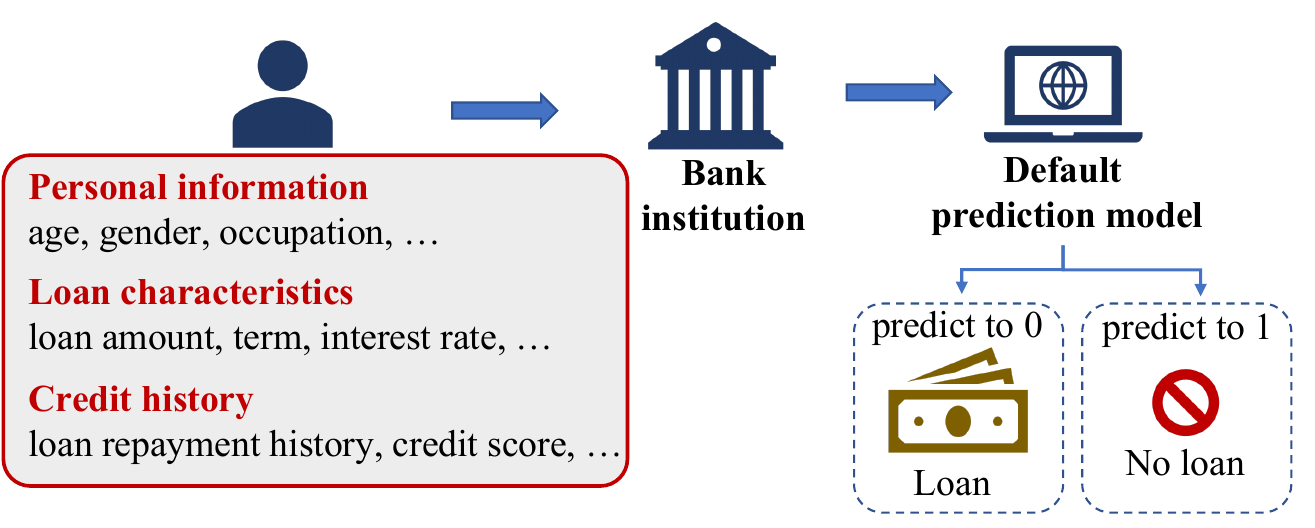}}
    \caption{Credit approval process.} \label{figure1}
\end{figure}

Credit default prediction constitutes a vital aspect of risk assessment and management within the financial sector. Default prediction evaluates the likelihood of borrowers fulfilling their repayment obligations by considering various factors, such as personal information, credit history, and loan characteristics, shown in Figure \ref{figure1}. In this work, credit default prediction is framed as a binary classification task, distinguishing between default (label 1) or no default (label 0). By developing a robust credit default prediction model, financial institutions could make informed decisions regarding credit approvals based on estimating default probabilities.

\section{Methodologies}
The low correlation and imbalance in credit data make it challenging to predict defaults accurately. To focus on important information, we first organize multi-view feature spaces by selecting features based on various linear or nonlinear importance indicators. We then employ MLP to obtain predictions from these local views. Simultaneously, for the global view of all features, we obtain a global feature space using MLP. To ensemble key knowledge for robust credit predictions, we develop self-attention modules to stack generalizing these obtained local and global outputs. Furthermore, we set the $f1$ score as training monitor indicators to adjust the learning rate to assist the model in finding the optimal state for identifying defaulting users, addressing the challenges posed by data imbalance. The code can be found in \footnote{https://github.com/jinxtan/Tab\_Attention/}.

\subsection{Multi-view feature spaces}

\begin{figure}[t] 
    \centering
    \includegraphics[width=1\linewidth]{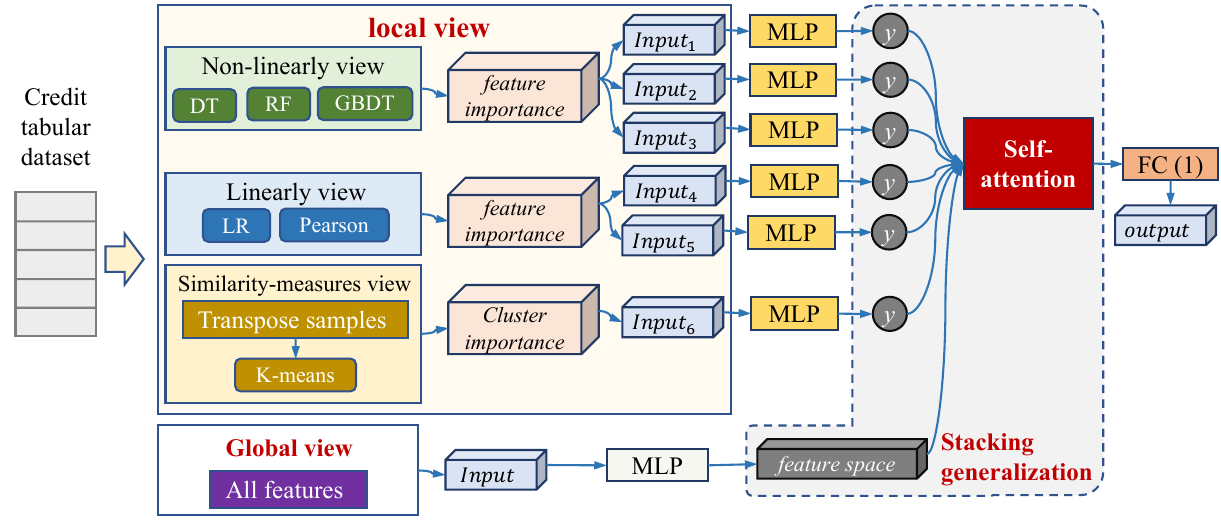}
    \caption{Multi-view feature spaces generation in Tab-Attention.}
    \label{figure2}
\end{figure}

In daily life, a general concept of "strong alliance" refers to the idea that by combining individuals with unique strengths, advantageous results can be obtained. Similarly, in feature selection, different approaches can lead to distinct advantageous views. To learn the multi-view information provided by these "strong alliance" feature sets, an automatic multi-view feature space generation method has been designed based on multiple screening strategies, shown in Figure \ref{figure2}. The following four types of views are considered:

\begin{itemize}
\item \textbf{Non-linearly correlated view}: This approach obtains the local-view feature spaces with the top 30$\%$ feature importance based on DT, GBDT, and RF methods. These models filter feature sets with non-linearly correlated importance based on information gain, which is important for default prediction models to focus on non-linear mappings between labels and features.

\setlength{\parindent}{1em} For $D =\{x_1,x_2,...,x_n,y\}$, we give the example that calculates feature importance based on a DT using $ID_3$:

\begin{enumerate}
\item Calculate the entropy of the target variable $y$ as:
\begin{equation}
    \Gamma(y) = - \sum (\Theta(y) / l) * \Theta(\Theta(y) / l),
\end{equation}

where $\Theta(y)$ refers to the number of occurrences of default and non-default samples in $y$, and $l$ is the total number of instances.
        
\item For each feature $x_i$, we calculate the information gain $x_{i_{IG}}$ as:
\begin{equation}
    x_{i_{IG}} = \Gamma(y) - \sum \left(\frac{\Theta(x_i)}{l}\right) \cdot \Gamma(y | x_i),
\end{equation}
where $\Gamma(y | x_i)$ is the entropy of $y$ given feature $x_i$.

\item Select the feature with the highest $x_{i_{IG}}$ as the current division node.

\item Create a branch for each possible value of the selected $x_i$ and recursively repeat steps 1-3 on the subsets of data corresponding to each branch.

\item Stop the recursion when:
\begin{enumerate}
\item All samples belong to the same class.
\item There are no more features to select.
\item The maximum depth of the tree is reached.
\item No enough instances could be split into a node.
\end{enumerate}

\item Assign the majority class of the instances in the current subset as the class label for the leaf node.

\item Calculate the feature importance for $x_i$ by accumulating its information gain $x_{i_{IG}}$ in the process of constructing the DT.
\end{enumerate}

\item \textbf{Linearly correlated view}: Linear correlation helps the model learn the linear mapping capability between features and the targets. Pearson correlation coefficient and LR are designed to select the features with the top 30$\%$ linear correlation importance, forming two local views.

\setlength{\parindent}{1em} For $D =\{x_1,x_2,...,x_n,y\}$, the linear feature importance is calculated as,
\begin{enumerate}
    \item For LR, the optimal model is obtained as,
    \begin{equation}
        y_{pred} = \beta_0 + \beta_1\cdot x_1 + \beta_2\cdot x_2 + ... + \beta_n\cdot x_n,
    \end{equation}
where $\beta_0$, $\beta_1$,..., $\beta_n$ are the coefficients for each feature. We select the top 30$\%$ of features based on the absolute value of \textbf{$\beta$}.
\item For Pearson method, we obtain the correlation coefficient $\iota_i$ for feature $x_i$ as,
\begin{equation}
    \iota_i = cov(x_i, y) / (\sigma_{x_i} * \sigma_y),
\end{equation}
where $cov(x_i, y)$ is the covariance between $x_i$ and y, $\sigma$ means the standard deviation. We also select the features based on the absolute of $\iota_i$.
\end{enumerate}

\item \textbf{Similarity-measures view}: We employ the K-means to measure the similarity among features, then selects the features with the cluster with the largest number of features as a local view. Here, $K=4$ is set to obtain multiple similarity feature sets that help the model learn information from a specific perspective rather than global information. The calculation process is as,
\begin{enumerate}
    \item Transpose $D =\{x_1,x_2,...,x_n,y\}$.
    \item Utilize the K-means to partition the features into four clusters.
    \item Enumerate the number of features within each cluster.
    \item Select the cluster with the maximum count of features as the similarity-measures view.
\end{enumerate}

\item \textbf{Global view}: This view contains all features and offers the possibility of global optimization.
\end{itemize}

For each local view, MLPs are then designed to capture various possibilities. The global view aims to obtain a global feature space, while the local views aim to learn local knowledge. Here the MLPs model structure of the global view is larger than that of the local view.

\subsection{Self-attention-based stacked generalization}
\begin{figure*}
    \centering
    \includegraphics[width=0.6\linewidth]{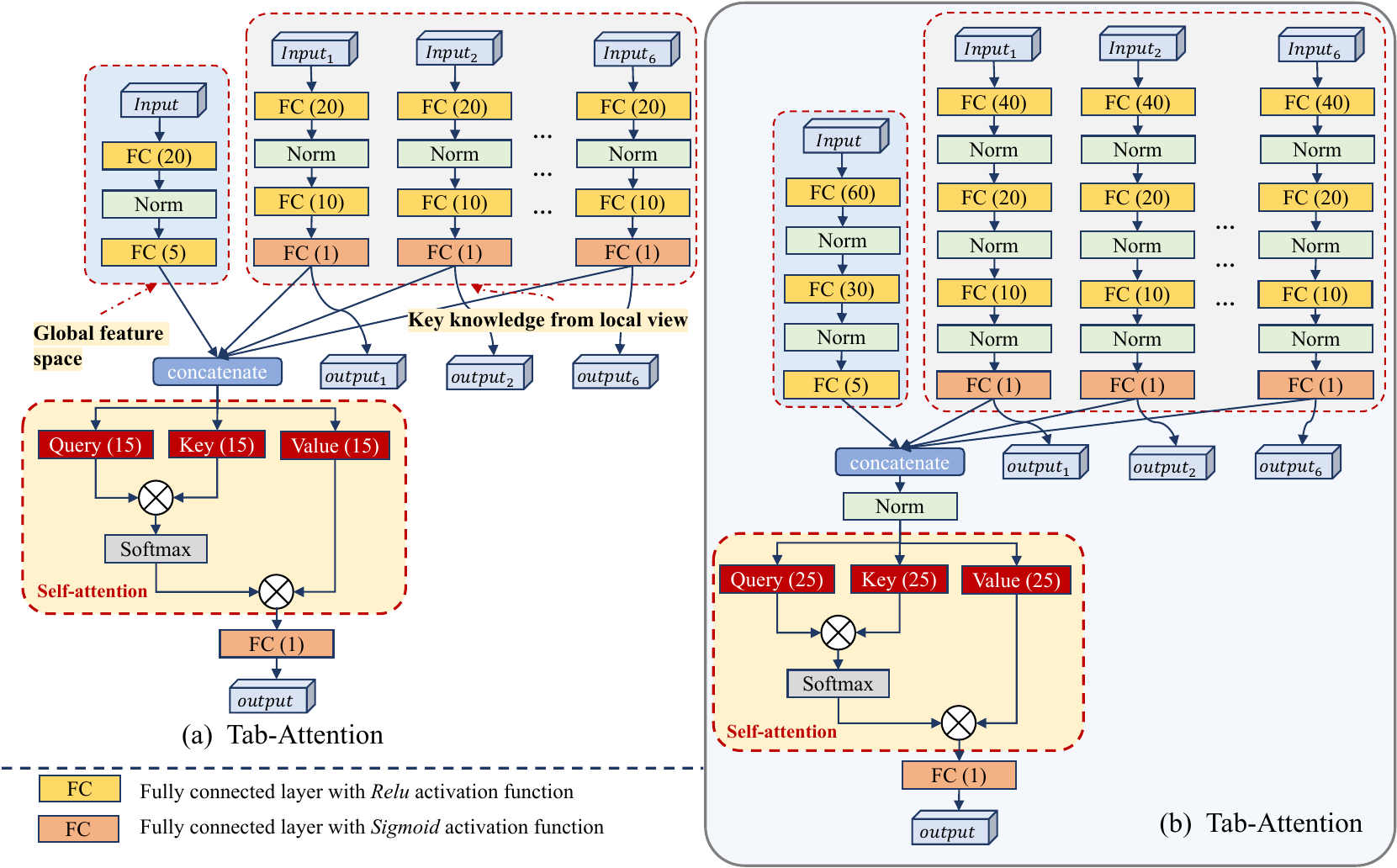}
    \caption{Tab-Attention model structure. (a) and (b) are the model structures of Tab-Attention for smaller and larger data sizes. The model described in the article is the structure of (a).}
    \label{figure3}
\end{figure*}

Compared to tabular data, financial data typically exhibits lower feature correlations, higher sparsity, and greater imbalance, making it more challenging to accurately identify default users. Actually, uncovering critical knowledge is essential for enhancing predictive performance. Attention mechanisms, which focus on essential data based on activation significance, have led to significant breakthroughs in computer vision \cite{RN109,bertasius2021space} and natural language processing \cite{niu2021review,dai2021coatnet}. Self-attention, a specific case of attention, learns the attention mechanism from the information provided by the data itself.

To address these challenges, we propose self-attention-based Tab-Attention to stack the knowledge from different feature views, as shown in Figure \ref{figure3}. Firstly, for different feature views $F=\{{input}_G,{input}_1,{input}_2,\ldots,{input}_6\}$ obtained in Section 3.1, we employ MLPs to acquire knowledge. The result is then normalized to reduce the risk of overfitting, as:

\begin{equation}
F^1=Norm(Relu(W_1\ast F+b_1\ )).
\label{thesystem}
\end{equation}

For the global view, we then also connect with the fully connected layer with fewer neurons to extract the global feature spaces as:
\begin{equation}
    F_G^2=Relu\left(W_2\ast F_G^1+b_2\right).
\end{equation}

For the local views, we also connect a fully connected layer and then feed it into the output layer to obtain the predictions as:
\begin{equation}
    F_i^3=Sigmoid\left(W_3\ast Relu\left(W_2\ast F_G^1+b_2\right)+b_3\right).
\end{equation}
Here, we optimize the cross-entropy between real labels and $F_i^3$ to learn the exclusive nonlinear or linear mapping knowledge from the local view. Next, we concatenate the acquired local knowledge $F_i^3$ and global feature space $F_G^2$ into the self-attention layer to learn the importance of different elements, where the concatenated vectors $F^*$ are linearly mapped to query matrix $Q$, key matrix $K$, and value matrix $V$, computed as,
\begin{equation}
    F^*=concat(F_G^2,F_1^3,F_2^3,…,F_6^3),
\end{equation}
\begin{equation}
    Q=W_Q*F^*+b_Q,
\end{equation}
\begin{equation}
    K=W_K*F^*+b_K,
\end{equation}
\begin{equation}
    V=W_V*F^*+b_V.
\end{equation}

Then, we calculate the attention value $\alpha$ between every two input vectors as:
\begin{equation}
    A=(\alpha)_{(i,j)}=softmax(Q*K).
\end{equation}
Based on the attention matrix $A$, we calculate the output vector $F^{attention}$ of the self-attention layer corresponding to each input vector $\alpha$ as,
\begin{equation}
    F^{attention}=A*V.
\end{equation}
Note that each element of $F^{attention}$ contains the associated information of other elements in $F^*$ so as to capture a wider compositional mechanism. Finally, we employ a fully connected layer to learn the mapping relationship to the real labels.
\begin{equation}
    y^{predict}=sigmoid(W_4*F^{attention}+b_4).
\end{equation}

For larger and smaller datasets, we configure different Tab-Attention model structures, as shown in Figure \ref{figure3}. Compared to smaller datasets, the Tab-Attention for larger ones involves a more complex model structure and normalization in all layers except for the output layer.

\subsection{Imbalanced learning based on $f1$ score}
Imbalanced credit data presents a significant challenge in default prediction. Traditional over-sampling \cite{sun2018imbalanced} and under-sampling \cite{itoo2021comparison} methods are prone to cause overfitting or underfitting. Additionally, the performance of weighted cross-entropy methods \cite{akil2020fully} is limited, while generative model-based methods \cite{fan2022data} are difficult to train and exhibit high complexity.

Traditionally, stopping training neural networks often relies on the convergence of overall loss or overall accuracy. However, this approach can inadvertently cause the model to neglect minority class samples. The $f1$ score, as the harmonic mean of $Precision$ and $Recall$, takes into account the prediction accuracy of default samples, calculated as,
\begin{equation}
    f1=2*(Precision*Recall)/(Precision+Recall),
\end{equation}
where $Precision = TP / (TP + FP)$, $Recall = TP / (TP + FN)$. The $TP$ and $FP$ are the default samples with default and non-default prediction labels, respectively. The $FN$ means non-default samples with default prediction labels. 

To address the imbalance challenges, we optimize the training process of Tab-Attention based on $f1$ score as a training monitoring indicator.  Specifically, when the $f1$ does not improve further after 10 epochs, we reduce the learning rate to 10$\%$ of the previous value, enabling the model to continue local optimization based on the current optimal state. The specific optimization process of imbalanced training is shown in Algorithm \ref{algorithm1}. Compared to the optimization process of traditional neural network training, this strategy facilitates the model to converge more rapidly to its optimal state of identifying default users.

\begin{algorithm}[!t]
  \caption{Training Tab-Attention based on $f1$ score.}
  \label{algorithm1}
  \begin{algorithmic}[1]
   \REQUIRE~~\\ The initialized weights $W$, $b$ and learning rate $\xi_0$ of neural networks, the best $f1$ score $f1_{best}=0$, and $epoch_{count}=0$. \\
        \FOR{each training epoch $i$}
        \STATE Train Tab-Attention on training set using the Adam optimizer with $\xi_0$.
        \IF{$f1>f1_{best}$}
        \STATE $ f1_{best}=f1$  \\
        \STATE $epoch_{count} = epoch_{count} + 1$.
        \ELSE
        \STATE $epoch_{count} = 0$. \\
        \ENDIF
        \IF{$epoch_{count} = 10$}
        \STATE $ \xi_{i+1}=\xi_i\ast0.1 $. \\
        \STATE Update the learning rate in the Adam optimizer.
        \ENDIF
        \IF{$epoch_{count} = 20$}
        \STATE Break the loop.
        \ENDIF
        \ENDFOR
   {\STATE \textbf{Output} Tab-Attention.}
  \end{algorithmic}
 \end{algorithm}

\section{Experimental results and analysis}

\subsection{Dataset and preprocessing}
\begin{table}
\renewcommand\tabularxcolumn[1]{m{#1}} 
\renewcommand{\arraystretch}{1.0} 
\centering
\caption{Statistics for all credit default datasets.}
\label{Table1}
\begin{tabularx}{\linewidth}{>{\centering\arraybackslash}X>{\centering\arraybackslash}X>{\centering\arraybackslash}X>{\centering\arraybackslash}X>{\centering\arraybackslash}X>{\centering\arraybackslash}X}
\hline
 &  &  & \multicolumn{2}{c}{$PCQ$} &  \\
Dataset & Data Size & Feature Num & 25\% & 75\% & $IR$ \\
\hline
Zhongyuan & 10,000 & 34 & 0.011 & 0.064 & 1:4.94 \\
Taiwan& 30,000 & 23 & 0.018 & 0.191 & 1:3.52 \\
South German & 1,000 & 20 & 0.033 & 0.155 & 1:2.33 \\
Statlog & 1,000 & 20 & 0.033 & 0.143 & 1:2.33 \\
LC2018 & 197,178 & 84 & 0.015 & 0.071 & 1:3.18 \\
LC2017 & 314,368 & 84 & 0.012 & 0.072 & 1:3.75 \\
\hline
\end{tabularx}
\end{table}

We utilized six credit datasets to validate the effectiveness of our model, shown in Table \ref{Table1}. We examined the data imbalance rate ($IR$) and the Pearson correlation quantiles ($PCQ$) between the features and labels, identifying challenges related to both imbalance and low correlation. We removed columns with ID attributes and over 60$\%$ missing values and imputed the remaining missing values with the mode. The target variable in these datasets is the default status (1 for default and 0 for non-default).
\begin{itemize}
    \item \textbf{Zhongyuan credit dataset}\footnote{https://www.datafountain.cn/competitions/530/datasets} (10,000 records, 34 features): from the 2021 CCF competition, focusing on individual credit default records and including loan records, user information, occupation, age, and marital status.
    \item \textbf{Taiwan credit dataset}\footnote{https://archive.ics.uci.edu/ml/datasets/default+of+credit+card+clients} (30,000 samples, 23 attributes): donated by Chung Hua University and Tamkang University in 2016, including user information, education level, repayment records, and credit card bills.
    \item \textbf{South German credit dataset}\footnote{https://archive.ics.uci.edu/ml/datasets/South+German+Credit} (1,000 records, 20 features): donated by Beuth University of Applied Sciences Berlin in 2019, including user information, loan amount, term, purpose, collateral, existing assets, and credit history.
    \item \textbf{Statlog credit dataset}\footnote{https://archive.ics.uci.edu/ml/datasets/Statlog+\%28German+Credit+Data\%29} (1,000 records, 20 features):  donated in 1994, including loan-related information, borrower personal information, and credit history.
    \item \textbf{LendingClub credit dataset}\footnote{https://www.lendingclub.com} (more than 197,178 samples, 84 attributes): containing data from the lending history of the US online lending platform LendingClub, including borrower personal information, loan information, credit rating and history, and loan purpose. We used data from 2018 and 2017.
\end{itemize}

Table \ref{Table1} presents that the IRs of the utilized datasets exceed 1:2, indicating data imbalance. Furthermore, the 75$\%$ $PCQ$ between the features and labels are below 0.2, indicating a weak correlation.

\subsection{Evaluation metrics}
Default prediction aims to minimize losses for financial institutions by denying loans to potential defaulters while ensuring access to creditworthy individuals. Therefore, we evaluate the model based on overall performance and class-specific performance:
\begin{itemize}
    \item \textbf{Overall performance}: We evaluate accuracy ($Acc$), area under the curve ($AUC$), and Kolmogorov-Smirnov ($KS$). $Acc$ measures the total number of samples classified correctly. $AUC$ can comprehensively assess model performance with the class-imbalanced dataset. $KS$, a widely used evaluation metric in the financial domain, measures the risk discriminatory ability by calculating the maximum difference between cumulative positive and negative instances, calculated as,
\begin{equation}
    KS=max(TPR-FPR),
\end{equation}
where $TPR = TP/(TP + FN)$ and $FPR = FP/(FP + TN)$ are the cumulative results of two class samples, respectively. 

\item \textbf{Class-specific performance}: for defaulting and non-defaulting samples, we assessed the $Recall$, $Precision$, and $f1$ score to analyze the prediction capability for two classes.
\end{itemize}

\subsection{Experimental setting}
The experiments were conducted on an Ubuntu 20.04 server with a 5.8 Linux kernel, featuring Matrox G200eW3 GPU drivers operating at 3.90 GHz and 16 GB of main memory. In this study, each model configuration was executed 20 times, with the random state for splitting the data into training and testing sets set to values between 0 and 20 to mitigate the differences arising from data bias. 

For the LC datasets, the batch size is configured to 300, 100 for the Taiwan dataset, 30 for Statlog and South German datasets, and 50 for Zhongyuan dataset. The initial learning rate for all datasets was set to 0.01. All data were normalized to minimize the impact of differences in scale.

\textbf{Baseline}: We compared our methods with three types of models: deep learning models (TabNet\cite{RN83}, NODE\cite{RN98}, and DNN), ensemble learning models (XGBoost\cite{RN87,RN95}, AdaBoost, GBDT\cite{RN91}, and RF\cite{RN90}), and machine learning-based single predictors (DT and LR \cite{nie2011credit}).

\textbf{Parameter configurations}: For TabNet, the number of neurons is set to 40 for the LC dataset, with 20 neurons for decision-making steps and 3 decision steps. For other datasets, the number of neurons is set to 20, with 10 neurons for decision-making steps. For the NODE model, we configured the number of neurons per layer to be 20 and employed 3 layers. Tree-based models have a maximum depth of 10, with 50 weak classifiers for ensemble models. The LR was optimized using L2 regularization with a regularization strength C of 0.1.

\subsection{Performance comparison}

\begin{figure*}
    \centering   
    \includegraphics[width=1\linewidth]{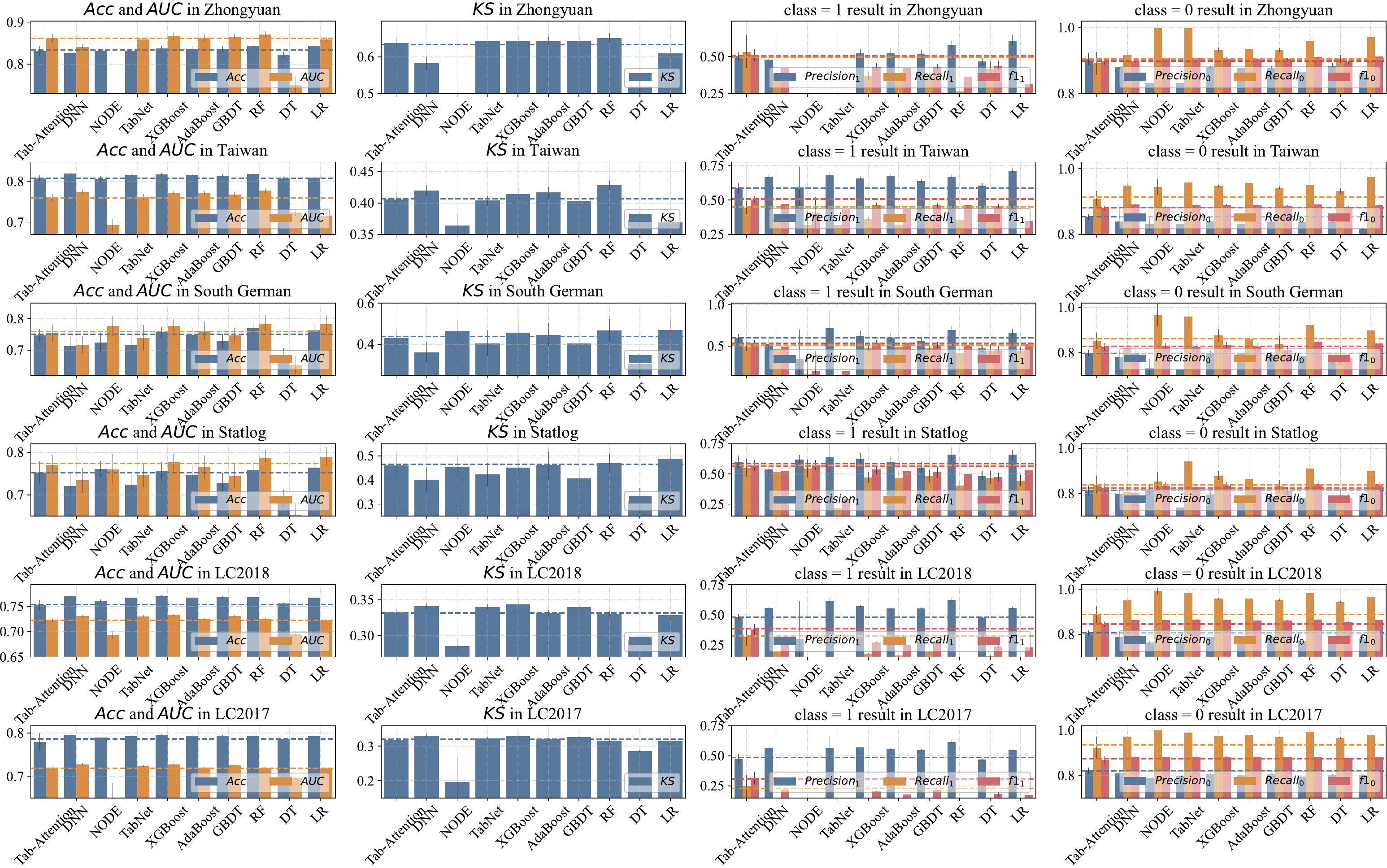}
    \caption{Performance comparison of Tab-Attention and current advanced models.}
    \label{figure4}
\end{figure*}

Figure \ref{figure4} displays the comparison results of model performance, and the analysis is as,
\begin{itemize}
    \item For the Zhongyuan dataset, Tab-Attention and ensemble learning models (XGBoost, AdaBoost, GBDT, RF) achieve similar overall advantageous performance ($Acc, AUC, KS$). In terms of primary default prediction tasks, TabNet and NODE fail to recognize defaulting users, whereas Tab-Attention reaches a $Recall_1$ of 53.3$\%$ for default samples, outperforming other approaches by 28.7$\%$, while maintaining outstanding $Precision_1$ of 50.4$\%$. Furthermore, Tab-Attention attains $f1_1 = 51\%$, which is 16.4$\%$ better than other models. 
    \item For the Taiwan dataset, similarly, Tab-Attention's $Recall_1 = 44.5\%$ surpasses other approaches by 20.9$\%$, with a $f1_1$ of 50.4$\%$, which is 7.2$\%$ better than other approaches. At the same time, Tab-Attention's overall performance and non-default sample prediction performance remain at a comparable level to the better results of other models. 
    \item For the South German and Statlog datasets with smaller sample sizes, Tab-Attention could also recall more default users while maintaining a superior precision ($Precision_1 \approx 0.6$), which also upholds an advantageous level for non-default user prediction and overall performance.
    \item For the LC dataset, Tab-Attention's $Recall_1$ result exceeds other models by more than 65.8$\%$, and its $f1_1$ score is over 31.7$\%$ better than other schemes. Meanwhile, the overall performance of Tab-Attention is at a comparably superior level to DNN, TabNet, XGBoost, AdaBoost, GBDT, RF, and LR.
\end{itemize}

In general, compared to existing advanced methods, Tab-Attention achieves superior default prediction performance ($Recall_1$ and $f1_1$ are 32.92$\%$ and 16.05$\%$ better than other models on average, respectively), while maintaining outstanding overall performance and $Recall_0$ and $Precision_0$ for non-default users. Noteworthily, Tab-Attention's $Recall_0$ is lower than other models, but its $Precision_0$ is superior, implying that Tab-Attention is less likely to mistakenly predict default users as non-default ones with massive samples.

\subsection{Ablation study}
\begin{figure*}
    \centering
    \includegraphics[width=1\linewidth]{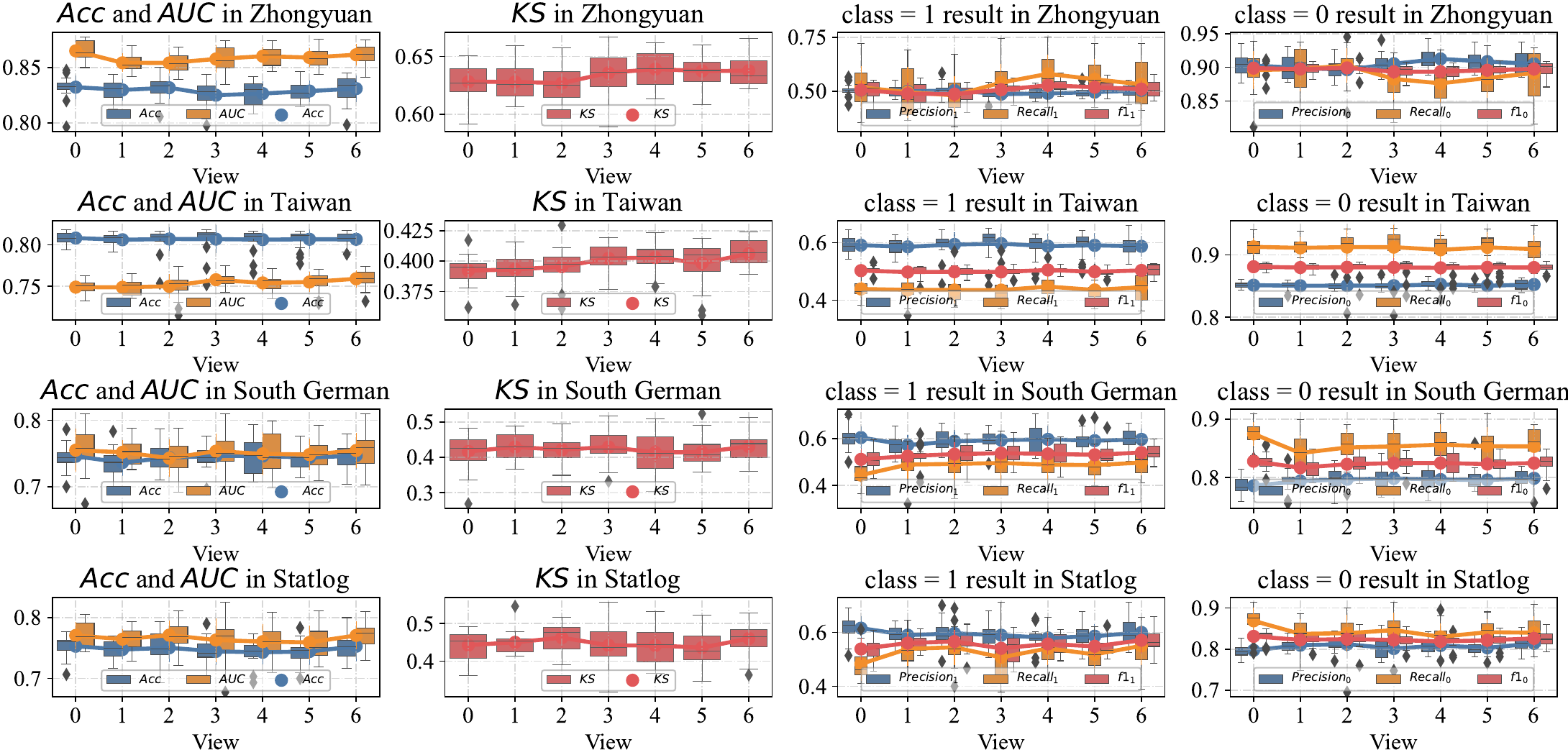}
    \caption{Results as the view increases. $View = 6$ means Tab-Attention, and $View = 0$ means DNN without the local-view model. 1, 2, 3, 4, 5, and 6 represent increasing views based on DT, GBDT, RF, LR, K-means, and Pearson, respectively.}
    \label{figure5}
\end{figure*}
We develop multi-view feature spaces and self-attention stacking generalization components to assist the model in mining key knowledge for more robust default predictions. In addition, we designed f1-based imbalance learning. To verify the effectiveness of these components, we performed the following ablation study.

\textbf{Multi-view feature space}: Figure \ref{figure5} shows the comparison results as more views feature spaces participate in the collaboration. As more views are incorporated, the $KS$ progressively improves. Compared to Tab-Attention ($View=0$) without multi-view feature spaces, the $KS$ of Tab-Attention increases by 1.6$\%$, 3.6$\%$, 3.4$\%$, and 3.8$\%$ for the Zhongyuan, Taiwan, South German, and Statlog datasets, respectively, while maintaining comparable levels of $Acc$ and $AUC$. For the primary task of default identification, compared to Tab-Attention ($View=0$), $Recall_1$ of Tab-Attention increases by 2.7$\%$, 1.4$\%$, 12.2$\%$, and 13.9$\%$ for the four datasets, respectively. $Precision_1$ remains roughly equivalent level, and $f1_1$ displays an increasing trend, with the addition of more views. Additionally, $Recall_0$ gradually decreases as the views increase, while $Precision_0$ presents an increasing trend, which indicates that the addition of different views can reduce the risk of the model overlooking minority default users. 

Overall, the cooperation of various views contributes to the model's ability to identify more default users and enhances the distinction between the two types of samples, resulting in the gradual improvement of the $KS$.

\begin{table}
\centering
\caption{Ablation comparison of f1-based imbalanced training and self-attention-based stacked strategies.}
\label{Table2}
\resizebox{\linewidth}{!}{
\begin{threeparttable}
\begin{tabular}{lcccc>{\columncolor{gray}}c}
\toprule
\multirow{11}{*}{\makecell[c]{Zhong \\ yuan}}
&Train metric & $f1$ & $AUC$ & $Acc$ & Tab* \\
\midrule
&$Acc$ & $0.83 \pm 0.01$ & \bm{$0.84 \pm 0.00$} & \bm{$0.84 \pm 0.00$} &$0.84 \pm  0.00$ \\
&$AUC$ & $0.86 \pm 0.00$ & $0.86 \pm 0.00$ & \bm{$0.87 \pm 0.00$} &$0.87 \pm  0.00$\\
&$KS$ &\bm{ $0.64 \pm 0.01$ }& \bm{$0.64 \pm 0.01$} & $0.62 \pm 0.01$ & $0.64 \pm 0.01$ \\
&$Precision_0$ &\bm{ $0.91 \pm 0.01$} & $0.89 \pm 0.00$ & $0.89 \pm 0.00$ & $0.89 \pm 0.01$ \\
&$Recall_0$ & $0.89 \pm 0.02$ & \bm{$0.93 \pm 0.01$} & $0.91 \pm 0.01$ & $0.91 \pm 0.01$ \\
&$f1_0$ & $0.90 \pm 0.00$ & \bm{$0.91 \pm 0.00$} & $0.90 \pm 0.00$ & $0.90 \pm 0.00$ \\
&$Precision_1$ & $0.50 \pm 0.01$ & \bm{$0.53 \pm 0.01$} & $0.51 \pm 0.01$ & $0.51 \pm 0.01$ \\
&$Recall_1$ & \bm{$0.53 \pm 0.05$} & $0.41 \pm 0.02$ & $0.45 \pm 0.02$ & $0.47 \pm 0.04$ \\
&$f1_1$ & \bm{$0.51 \pm 0.02$} & $0.46 \pm 0.01$ & $0.48 \pm 0.01$ & $0.48 \pm 0.02$ \\
&$epoch$ & \bm{$38 \pm 5$} & $77 \pm 6$ & $95 \pm 5$ & $40 \pm 7$ \\
\midrule
\multirow{10}{*}{Taiwan}
&$Acc$&$0.81\pm0.00$&\bm{$0.82\pm0.00$}&\bm{$0.82\pm0.00$}&$0.81\pm0.00$\\
&$AUC$ &$ 0.76\pm0.00$&\bm{$0.77\pm0.00$}&\bm{$0.77\pm0.00$}&$0.76\pm0.01$\\
&$KS$ &$0.41\pm0.01$&\bm{$0.42\pm0.00$}&\bm{$0.42\pm0.00$}&$0.41\pm0.01$\\
&$Precision_0$&\bm{$0.85\pm0.00$}&$0.84\pm0.00$&$0.84\pm0.00$&$0.85\pm0.00$\\
&$Recall_0$&$0.91\pm0.01$&\bm{$0.95\pm0.00$}&\bm{$0.95\pm0.00$}&$0.92\pm0.01$\\
&$f1_0$ &$0.88\pm0.00$&\bm{$0.89\pm0.00$}&\bm{$0.89\pm0.00$}&$0.89\pm0.00$\\
&$Precision_1$ &$0.59\pm0.02$&\bm{$0.67\pm0.01$}&\bm{$0.67\pm0.01$}&$0.61\pm0.01$\\
&$Recall_1$ &\bm{$ 0.45\pm0.02$}&$0.36\pm0.01$&$0.36\pm0.01$&$0.42\pm0.01$\\
&$f1_1$ &\bm{$0.50\pm0.01$}&$0.46\pm0.01$&$0.47\pm0.01$&$0.50\pm0.01$\\
&$epoch$ &\bm{$31\pm1$}&$66\pm6$&$64\pm9$&$32\pm1$\\
\midrule
\multirow{10}{*}{\makecell[c]{South \\ German}}
&$Acc$&\bm{$0.75\pm0.01$}&$0.74\pm0.01$&$0.73\pm0.01$&$0.75\pm0.01$\\
&$AUC$&\bm{$ 0.75\pm0.01$}&\bm{$0.75\pm0.01$}&$0.71\pm0.01$&$0.75\pm0.01$\\
&$KS$ &\bm{$0.43\pm0.02$}&$0.42\pm0.02$&$0.37\pm0.02$&$0.42\pm0.03$\\
&$Precision_0$&\bm{$0.80\pm0.01$}&$0.79\pm0.01$&$0.79\pm0.01$&$0.79\pm0.01$\\
&$Recall_0$&$0.85\pm0.02$&\bm{$0.86\pm0.02$}&$0.83\pm0.01$&$0.87\pm0.01$\\
&$f1_0$ &\bm{$0.82\pm0.01$}&\bm{$0.82\pm0.01$}&$0.81\pm0.01$&$0.83\pm0.01$\\
&$Precision_1$ &\bm{$0.60\pm0.02$}&$0.59\pm0.03$&$0.55\pm0.02$&$0.60\pm0.03$\\
&$Recall_1$&\bm{$0.50\pm0.02$}&$0.49\pm0.02$&$0.49\pm0.03$&$0.47\pm0.02$\\
&$f1_1$ &\bm{$0.54\pm0.02$}&$0.53\pm0.01$&$0.51\pm0.02$&$0.52\pm0.02$\\
&$epoch$ &\bm{$56\pm6$}&$74\pm5$&$121\pm12$&$59\pm5$\\

\midrule
\multirow{10}{*}{Statlog}
&$Acc$&\bm{$0.75\pm0.01$}&$0.74\pm0.01$&$0.73\pm0.01$&$0.75\pm0.01$\\
&$AUC$ &\bm{$ 0.77\pm0.01$}&\bm{$0.77\pm0.01$}&$0.73\pm0.01$&$0.77\pm0.01$\\
&$KS$ &\bm{$0.46\pm0.02$}&\bm{$0.46\pm0.02$}&$0.40\pm0.03$&$0.45\pm0.02$\\
&$Precision_0$&\bm{$0.81\pm0.01$}&\bm{$0.81\pm0.01$}&$0.8\pm0.01$&$0.80\pm0.01$\\
&$Recall_0$&\bm{$0.84\pm0.02$}&$0.83\pm0.01$&$0.82\pm0.01$&$0.85\pm0.01$\\
&$f1_0$ &\bm{$0.83\pm0.01$}&$0.82\pm0.01$&$0.81\pm0.01$&$0.83\pm0.01$\\
&$Precision_1$ &\bm{$0.60\pm0.02$}&$0.58\pm0.02$&$0.55\pm0.02$&$0.60\pm0.02$\\
&$Recall_1$ &\bm{$0.55\pm0.03$}&$0.54\pm0.03$&$0.52\pm0.03$&$0.51\pm0.03$\\
&$f1_1$ &\bm{$0.57\pm0.02$}&$0.56\pm0.02$&$0.53\pm0.02$&$0.55\pm0.02$\\
&$epoch$ &\bm{$56\pm5$}&$76\pm7$&$117\pm10$&$54\pm5$\\
\bottomrule
\end{tabular}
     \begin{tablenotes}    
         \footnotesize               
         \item[1] Tab* means Tab-Attention without self-attention.
    \end{tablenotes}           
\end{threeparttable} }
\end{table}

\textbf{Self-attention-based stacked generalization}: Then, we validated the necessity of ensembling important knowledge using the self-attention mechanism, as shown in Table \ref{Table2}. Overall, for the four datasets, the inclusion of self-attention increased the $Recall_1$ of default samples by more than 5$\%$, and the $f1_1$ improved by 3.45$\%$ on average. Furthermore, Tab-Attention shows comparable overall performance ($Acc$, $AUC$, and $KS$) and the prediction results for non-default samples, compared to Tab*. This demonstrates that the self-attention mechanism is capable of ensembling essential knowledge to help the model better predict default users. 

\textbf{Imbalanced learning based on $f1$ score}: Lastly, to validate the necessity of $f1$ in imbalanced learning, we compared it with $AUC$ and $Acc$ as monitoring indicators for the learning rate adjustment, shown in Table \ref{Table2}. Firstly, for the Zhongyuan dataset, it can be observed that the $Recall_1$ and $f1_1$ of the $f1$ training scheme are more than 17.8$\%$ and 6.3$\%$ better than that of $Acc$ and $AUC$, respectively, while maintaining an excellent $Precision_1 = 0.5$. Meanwhile, $Recall_0$ of $Acc$ and $AUC$ training schemes is significantly higher than that of $f1$, but their $Precision_0$ is lower than that of $f1$, indicating that $Acc$ and $AUC$ schemes are more likely to misclassify default samples as normal samples. In addition, there is no significant difference in the $Acc$, $AUC$, and $KS$ results among the three training schemes. For the Taiwan dataset, similarly, the $Recall_1$ and $f1_1$ of the $f1$ scheme are 25$\%$  and 6.4$\%$ better than other schemes, respectively. Furthermore, $Acc$ and $AUC$ schemes also expose the risk of recalling more normal samples as default samples (with superior $Recall_0$ but inferior $Precision_0$). For the South German and Statlog datasets, the performance of $f1$ in various indicators  is significantly superior to that of $f1$ and $AUC$. Moreover, $f1$ imbalance learning converges more than 32.1$\%$ faster than the other two strategies, i.e. with fewer epochs. 

Overall, the $f1$ training strategy is beneficial for Tab-Attention to accurately identify default users while maintaining superior overall $Acc$, $AUC$, $KS$, and non-default user prediction performance. 

\section{Conclusion}

In this work, to address the imbalanced and low feature-related credit default prediction, we proposed Tab-Attention, a novel credit default prediction framework that stacks important knowledge from multi-view feature spaces. Compared with the existing advanced GBDT-based and deep learning-based models, Tab-Attention improves the $Recall_1$ and $f1_1$ of default users by about 32.92$\%$ and 16.05$\%$ on average, respectively. We also observed that the collaboration of multi-view feature spaces helps Tab-Attention to better identify default risk, with an increasing trend of $KS$, $Recall_1$ and $f1_1$. In addition, the addition of self-attention is able to better ensemble important knowledge to further achieve superior $Recall_1$ and $f1_1$. Furthermore, we employed an $f1$-based imbalanced training strategy to assist the model to converge faster to the optimal state for identifying defaulting users. In future work, we would derive a theoretical justification on why Tab-Attention can better identify default users.

\ack We would like to thank the referees for their comments, which
helped improve this paper considerably. The work is supported by the National Key Research and Development Program of China (2021YFC3300600). 

\bibliography{ecai}

\end{document}